# AI-Mediated Explainable Regulation for Justice

___________________________________________________________________

Thomas Hofweber[1]
University North Carolina
at Chapel Hill

Andreas Sudmann
University of Bonn

Evangelos Pournaras
University of Leeds

**Abstract**

Present practice of deciding on regulation faces numerous problems that make adopted regulations static, unexplained, unduly influenced by powerful interest groups, and stained with a perception of illegitimacy. These well-known problems with the regulatory process can lead to injustice and have substantial negative effects on society and democracy. We discuss a new approach that utilizes distributed artificial intelligence (AI) to make a regulatory recommendation that is explainable and adaptable by design. We outline the main components of a system that can implement this approach and show how it would resolve the problems with the present regulatory system. This approach models and reasons about stakeholder preferences with separate preference models, while it aggregates these preferences in a value sensitive way. Such recommendations can be updated due to changes in facts or in values and are inherently explainable. We suggest how stakeholders can make their preferences known to the system and how they can verify whether they were properly considered in the regulatory decision. The resulting system promises to support regulatory justice, legitimacy, and compliance.

**Introduction**

Among the many aspects of government that affect the lives of its citizens, regulation has often been criticized both in the public debate and in scholarship as inefficient, opaque, and inflexible. While legislation gets adopted after public debate in parliament or, in the case of the United States, in Congress, regulation generally is decided upon with much less transparency. Consequently, it often remains obscure why certain regulations were adopted, which shields the reasons for adopting them from public scrutiny. This in turn leads to stasis and obstacles to changing existing regulation in light of changes in society, while allowing vulnerable groups to suffer injustice. These flaws in the present regulatory process are supported by decades of scholarship on regulatory capture and the influence of special interest groups on the regulatory process.[2] And they are illustrated by contemporary examples like Louisiana's 'Cancer Alley', where poorer populations are exposed to extensive carcinogenic air pollution despite many efforts at environmental regulation.[3]

---

[1] Corresponding author: hofweber@unc.edu
[2] Carpenter and Moss 2013.

In this article we argue that a value-sensitive socio-technical approach, informed by recent advances on trustworthy artificial intelligence (AI), can address this longstanding problem. We outline a vision and roadmap of an AI-mediated, explainable regulatory process, which is sensitive to changes in the world and the values of society, as well as subject to critical rational scrutiny. We argue that utilizing distributed AI can achieve explainable regulation that does not need to rely on inscrutable judgments by a single party. Instead, such a system can model the interests of various stakeholders and then aggregate them based on rigorous principles of social choice theory in a value-sensitive way. This approach explicitly highlights how different interests of stakeholders are weighted against each other in a way that is sensitive to what values one hopes to realize. Such a paradigm is not only subject to critical evaluation, but also automatically sensitive to any change in the values one hopes to realize or any change in the facts on which a regulatory decision was based. Such changes could automatically trigger a call to revise regulation and a suggestion for how such a revision should go. We will outline and discuss this vision for an improved regulatory process in this commentary and highlight its potential to foster justice and the public good.

**The opacity of the regulatory process**

As a first approximation, we can distinguish legislation from regulation. Legislation is the result of parliamentary debate and a public vote, which establishes a set of rules that govern society. However, often these rules are broad and need to be implemented with more detail. Regulation provides these details and specifies what precisely should be done to implement the law. The Clean Air Act[4] in the US, to consider one example, sets a general policy framework with certain goals and values: reduce pollution and improve public health. These policies are then implemented by the Environmental Protection Agency (EPA) in the regulatory process. This requires the EPA regulators to make tradeoffs between minimizing pollution levels and minimizing economic impact that results from restrictions on what is permissible. Such tradeoffs always reflect a balance of values: how much should we value risks to health and how much should we value risks to the economy? Whereas legislation happens in public view, regulation generally does not. It is often the result of decisions made by individual regulators away from public scrutiny. Regulators normally base their decisions on substantial research, but they are likely also exposed to lobbying attempts to influence their decision making by various stakeholders and interest groups. But even in the best of circumstances, where the regulator aims to implement a policy that reflects the known facts and aims to realize the values targeted by the legislation, while also trying to avoid being unduly influenced by interest groups, this process is subject to several flaws and limitations.

---

[3] Lerner 2010
[4] Clean Air Act, 42 U.S.C. § 7401 et seq.

First, the present regulatory process is not flexible enough to respond to a changing world. Once the regulatory decision has been made and the regulation is finalized it becomes static and requires considerable effort to motivate revisiting and changing it. Although some aspects of the regulatory decision making will have been preserved on paper, revisiting the issue will essentially require looking at all the facts anew and weighing all the considerations all over again. This is particularly true when societal values change and the proper target of the regulation should no longer be to achieve one value, but rather another one. How regulation should be changed considering this shift in value is complex and hard to determine other than by starting the regulatory process all over again, at substantial cost. This insensitivity of regulation to change also applies to changes in facts: new technologies can reduce certain harms, or new threats can increase exposure to certain risks. Scientific research might discover that a particular pollutant is more harmful than originally thought or changes in society might put a greater value on environmental protection than when environmental regulation was originally put in place. All these changes could justify a change in how a particular chemical is to be regulated. But to overcome regulatory inertia requires a coordinated effort and sufficient influence, something that favors more well-coordinated and well-funded interest groups over individuals, leading to ongoing regulatory injustice. Ideally, regulation should be sensitive to such changes by design, automatically flagging when regulatory decisions warrant revisiting due to changes in facts or in values.

Second, regulatory decisions are not easily explainable to members of the public. Even if the decisions are made for good reasons, those reasons are not generally known, justifiable, or appreciable by citizens. They might have been on the mind of the regulator when they made the decision, but those reasons are often not recoverable from the decision after it was made, and thus it is hard to cite those reasons to explain to members of the public why this regulatory decision was made. Decisions that cannot be supported by easily accessible good reasons can seem arbitrary and illegitimate, even when they were made with the best intentions and for good reasons. This is closely connected to the problem of stasis and lack of adoptability discussed just above. When the reasons for a decision are not transparent and traceable it is hard to evaluate whether those reasons are still sufficient to support the decision, whether the values on which it was based still hold, and whether a different decision might not be warranted today. And what's more, when the reasons are not available for public scrutiny there is the real danger that they would not withstand such scrutiny and that thus an unfair and unjust regulatory decision remains in place when it should not.

Third, the present regulatory process is subject to undue influence of more powerful stakeholders and can exclude marginalized groups. This is again illustrated by environmental regulation like that affecting `Cancer Alley' mentioned above.[5] Even when regulators are not unduly influenced by lobbyists, there is nonetheless an unequal amount of influence on the regulator's decision from more organized industry groups compared to individuals who are already disadvantaged and marginalized. Individuals need to organize a community group or raise issues in traditional media or on social media, but such efforts only succeed with a substantial investment of time, energy, and other resources, something marginalized groups have little of. This results in at least an unequal degree in how vivid the preferences of well-organized groups are compared to those that are marginalized, resulting in further disadvantage of the already disadvantaged, and thus more regulatory injustice.

---

[5] See Lerner 2010 for a more detailed analysis.

Fourth, the fact that regulation often happens behind closed doors with larger representation of those who have means and influence is widely known and affects the perception of part of the exercise of power of the government being dubious and illegitimate.[6] This in turn undermines the legitimacy of not just the regulatory decision, but the government's exercise of power more generally. It gives rise to the impression that although legislation might be achieved legitimately by elected officials after public debate, regulation is settled in private by appointed regulators who hear more from the powerful than the marginalized. And since regulatory decisions have a profound effect on the exercise of power of the government over its citizens, the perception of this exercise of power as legitimate is undermined by the present regulatory practices. This in turn negatively affects regulatory compliance. Citizens are more likely to comply with decisions they perceive as legitimate, and the imperfect regulatory process is thus a source of undermining the government's authority and legitimacy.

Present regulatory practices thus have quite some room for improvement. They can be inefficient and static, they often are not explainable to the people affected by the regulation, they sometimes further disadvantage the already disadvantaged, and they can undermine the legitimacy of the government. This problem has a profoundly negative effect on our democracy, leading not only to further injustice and suboptimal regulatory decisions, but also an undermining of regulatory compliance and legitimacy. In the following we will articulate a vision of how to address these problems. Although attempts to improve the regulatory process have generally failed,[7] we suggest that the introduction of value-sensitive distributed AI systems opens a new avenue for progress. With this, we envision a novel and socially responsible use of AI that can transform regulation while at the same time eliminating the risks inherent in using AI in the regulatory process.

**Reimagining the regulatory process with distributed AI**

The potential AI has for improving democratic and regulatory processes is currently challenged by the overwhelming risks that stem from the recent developments of AI, for instance, generative AI and large language models. We do not deny these risks but hope to make the case that there is also tremendous potential for breakthrough progress, especially in the field of regulation supported by alternative AI approaches, that are distributed and value sensitive by design. Existing AI systems have already been used for drafting and amending regulatory texts,[8] but this falls far short of the fundamental improvement to the regulatory process we envision, which aims to solve the problems listed above. Here AI should not be seen as a magic bullet that can simply solve this problem. In fact, handling regulatory decisions with existing AI systems would not help with these problems and likely make things worse. If we tasked a single AI model to simply make the regulatory decision based on a given legislation and various inputs from stakeholders, then this would not help with explainability. Such a model would likely be a large language model such as Claude or GPT, which are not only proprietary and black boxes to the public, they also do not reveal reasons for the decisions they recommend, and they can be subject to biases and as a result, further injustices. If a model is asked to present reasons for a decision, then it nonetheless remains unclear whether those reasons actually determined its decision or were merely an after-the-fact attempt to support it. A single, monolithic model would be even more inscrutable than a human regulator and relying on it would be a step in the wrong direction.

---

[6] See Fung and Weil 2007 and Freeman and Langbein 2000.
[7] See, for example, Beierle and Cayford 2002, as well as Kerwing and Furlong 2018.
[8] See Sanders and Schneider 2025.

What we need instead is a trustworthy distributed AI system grounded in social choice theory. Such a system will utilize different `agents' or agentic AI models representing different stakeholders in the regulatory decision process. Each such agent will model the preferences and interests of a particular stakeholder group, and all these agents together form a distributed multi-agent system.[9] Those agents then need to combine these preferences into a collective decision that determines regulation within the bounds of a given legislation. This essentially involves two parts: preference elicitation and preference aggregation.

Preference elicitation aims to capture the complex set of preferences that a stakeholder group might have with regards to a regulatory decision. The model will attempt to learn these preferences from stakeholder input. This might be a simple binary choice, a complex ranking of options, or some other preferential and expressive method to capture preferences. These preferences of the individual stakeholders then need to be aggregated into a collective decision. Such an aggregation will be based on a general method, algorithm or principle of aggregation. Utilizing different aggregation principles will express different values that are targeted in a collective decision. To illustrate how different aggregation principles realize different values, consider a decision problem where numerous polluting power plants need to be located either all in one neighborhood or else spread out more uniformly across different neighborhoods. If we aggregate preferences simply by a majority vote, then all those who do not live in the unfortunate neighborhood that gets all the power plants would prefer to have it there, winning the majority vote. If we aggregate votes more proportionally by how much a person would be negatively affected by the decision, then those who live in the unfortunate neighborhood would have greater influence on the decision, possibly enough to avoid the power plants being all placed in the same location. The former preference aggregation procedure favors the majority, while the latter, proportional representation by degree of negative effect, favors fairness, giving more voice to those who would be the worst off.[10] Many other options are on the table as well.

Preference aggregation is value sensitive in the sense that which one is the proper one to use depends on the values one hopes to realize in a collective decision problem. The distributed multi-agent system will make these values clear by explicitly basing collective decisions on one or another of these aggregation methods. This transparency will make a crucial difference in the resulting system being explicitly value sensitive and adoptive.

---

[9] See Shoham and Leyton-Brown 2009.
[10] See Parfit 1997, Scanlon 1975, Broome 1991, and Peters et al. 2021.

We envision such a multi-agent system to make a recommendation to a regulator about how a regulatory decision should be made. It would model stakeholder preferences and pick an aggregation method based on the value that it aims to realize and recommend a decision to a regulator based on this together with known facts and the law. The resulting recommendation can lay claim to being the right regulatory decision given the values one hopes to realize, the facts on the ground, and the preferences of the stakeholders.

Although multi-agent systems and distributed AI have been developed and applied in numerous cases, no such system has been built to help in the regulatory process. Developing this framework and building such a system would be of great benefit for society and democracy. To build such a system is not merely a problem for computer science or social choice theory. It involves many other aspects, in particular questions of value, fairness, justice, and how to achieve them in collective decision making and democracy. One aspect is to properly model the values and preferences of stakeholders, which requires getting input from them in a way that is feasible and informative. It will also require some thought on how to elicit preferences most accurately and informatively from the least amount of effort on behalf of the stakeholders. This will not be easy to implement but given the significance of the regulatory decision making it will be well worth the effort to try, since, as we will now argue, it would solve the problems with the regulatory process we discussed earlier.

**Resolving the problems with regulatory decision making**

We argued above that at present the regulatory process is flawed, leading to at least four serious problems that profoundly and negatively affect society. We can see that building and deploying a distributed AI system as outlined can be a key component in solving these problems and transforming the regulatory process for better and more just results.

First, the recommendation system would be straightforwardly adoptable to changing values and changing facts. It is value sensitive and makes the values it aims for explicit by deploying a particular method of aggregating preferences. If values change, the system can automatically reconsider the regulatory recommendation by aggregating differently or by using a different aggregation method altogether. If the facts change, the system can reaggregate based on the new facts. In either case, the recommendation can be quickly revisited at little cost. The recommendation system can be updated with new information to automatically revisit the regulatory recommendation and notify regulators that past decisions have lost their validity in light of these changes, and it can recommend a new regulatory decision.

Second, the recommended regulatory decision would be explainable. It would simply be the result of the selected aggregation method, aiming at the proper value, and the represented preferences of the stakeholder groups together with various facts about the world. There would be no mystery how the recommendation was arrived at. The regulator to whom the recommendation was made might follow it for just those reasons, or they might disagree with it and make a different decision. In that latter case it might be advisable, and could even be required, to give a justification for why the recommendation was not followed. This justification should be publicly available to make the recommendation again explainable and transparent. We do not envision simply outsourcing the regulatory process to AI but instead conceive of the multi-agent system as a tool that makes transparent recommendations to regulators.

Third, the regulatory process so imagined can easily accommodate input from affected stakeholders. We envision an online system or phone app where individuals affected by regulation can make their preferences known, and these preferences can then be represented in the agent that models the interests of this stakeholder group. The ideal implementation of this input mechanism needs to be worked out, but it can draw on existing platforms that were developed to aid direct democracy, e.g. CONSUL, DECIDIM, or the recent Stadtidee project in Aarau. The potential to directly influence a regulatory decision is thus straightforward within the distributed AI approach. It will become explicit whether and how the preferences of stakeholders were included in the final decision recommendation, not something to be taken on faith.

Fourth, the legitimacy of the regulatory decision will be substantially amplified by having an explanation available to citizens that is scrutable and subject to rational criticism. The multiagent system and the deployed aggregation method should be made publicly available. The justification and explanation for why the regulatory decision was made can be given explicitly, and the values that the decision aims for will be explicit as well and equally subject to critical scrutiny. This incorporates regulatory decision making in the democratic process, thereby gaining legitimacy and increasing compliance.

**Conclusion**

The present practice of regulatory decision making is flawed and has so far been resistant to substantial improvement. We argued that the regulatory process can be significantly augmented with a multi-agent AI system to model regulatory decision making in a value-sensitive way, leading to a regulatory recommendation. This system would be adoptable by design and lead to regulatory decisions that can be explained and justified by the very nature in which it came about. We need to study and develop such systems of AI-aided social decision making. This will require interdisciplinary collaboration, involving not just computer science and decision theory, but also law, media studies, political science, philosophy, and others. It promises to lead to substantial benefit to society, in the case of regulation by leading to better, more just regulatory decisions and by ensuring regulatory compliance and legitimacy via explainable decision making. AI can help achieve this, but not the monolithic, opaque, and biased models that are widely deployed today. We need to specifically develop explainable multi-agent models that individually represent different components of the decision problem, in particular stakeholders whose preferences are aggregated in an explicit, value-sensitive way, reflecting the ideal rational decision of each agent. The resulting regulatory recommendation would be explainable and adaptable to changes in facts and in values. It would be subject to rational criticism and thus regulation would be more properly incorporated into the public discourse, strengthening society and democracy.[11]

**References**

Beierle, T. C., & Cayford, J. (2002). *Democracy in Practice: Public Participation in Environmental Decisions*. Resources for the Future Press.

Broome, J. *Weighing Goods* (1991) Basil Blackwell.

Carpenter, D, & Moss, D.A. eds. (2013). *Preventing Regulatory Capture: Special Interest Influence and How to Limit It*. Cambridge University Press.

Fung, A., Graham, M., & Weil, D. (2007). *Full Disclosure: The Perils and Promise of Transparency*. Cambridge University Press.

Hänggli Fricker, R., Wellings, T., Zai, F., Yang, J.C., Majumdar, S., Bernhard, L., Weil, L., Hausladen, C.I. and Pournaras, E., (2024). Exploring legitimacy in a municipal budget decision in Switzerland: empirical insights into citizens' perceptions. *Philosophical Transactions A*, *382*(2285), p.20240098.

Hausladen, C.I., Hänggli Fricker, R., Helbing, D., Kunz, R., Wang, J. and Pournaras, E., (2024). How voting rules impact legitimacy. *Humanities and Social Sciences Communications*, *11*(1), pp.1-10.

Freeman, J., & Langbein, L. I. (2000). "Regulatory Negotiation and the Legitimacy Benefit." *NYU Environmental Law Journal*, 9, 60-151.

Kerwin, C. M., & Furlong, S. R. (2018). *Rulemaking: How Government Agencies Write Law and Make Policy* (5th ed.). CQ Press.

Lerner, Steve. (2010). *Sacrifice Zones: The Front Lines of Toxic Chemical Exposure in the United States* MIT Press

Parfit, D. (1997) "Equality or Priority?" In *The Ideal of Equality*, edited by Matthew Clayton and Andrew Williams, 81-125. Basingstoke: Palgrave Macmillan.

Peters, D., Pierczyński, G. and Skowron, P., (2021). Proportional participatory budgeting with additive utilities. Advances in Neural Information Processing Systems, 34, pp.12726-12737.

Pournaras, E., Majumdar, S., Wellings, T., Yang, J.C., Heravan, F.B., Fricker, R.H. and Helbing, D., (2025). Upgrading Democracies with Fairer Voting Methods. *arXiv preprint arXiv:2505.14349*

Sanders, N. & Schneider, B. (2025) "AI Will Write Complex Laws" *Lawfare* https://www.lawfaremedia.org/article/ai-will-write-complex-laws

---

[11] We are indebted to Sofia Ranchordás for many fruitful conversations.

Scanlon, T. (1975) "Preference and Urgency" Journal *of Philosophy* 72, no. 19: 655-669.

Shoham, Y. & Leyton-Brown, K. (2009) *Multiagent Systems: Algorithmic, Game-Theoretic, and Logical Foundations* Cambridge University Press